\documentclass[]{aa}

\usepackage{graphicx}
\usepackage{xcolor}
\usepackage{txfonts}
\usepackage{lipsum}
\usepackage{subcaption}         
\usepackage{lscape}             
\usepackage{placeins}           

\usepackage[hidelinks]{hyperref}

\begin{document}
    \title{Shared-optical-path VLBI frequency phase transfer from 86 to 258\,GHz on an 8600\,km baseline} \subtitle{Demonstrated with the APEX and IRAM 30\,m telescopes}


   \author{ 
G.-Y.~Zhao\inst{\ref{mpifr}},
A.~L.~Roy\inst{\ref{mpifr}},
J.~F.~Wagner\inst{\ref{mpifr}},
E.~Donoso\inst{\ref{concep}},
P.~Torne\inst{\ref{iram-es}},
E.~Ros\inst{\ref{mpifr}},
M.~Lindqvist\inst{\ref{oso}},
A.~P.~Lobanov\inst{\ref{mpifr}},
V.~Ramakrishnan\inst{\ref{turku}, \ref{aalto}},
T.~P.~Krichbaum\inst{\ref{mpifr}},
H.~Rottmann\inst{\ref{mpifr}},
J.~A.~Zensus\inst{\ref{mpifr}},
J.~P.~Pérez-Beaupuits\inst{\ref{eso-cl}, \ref{mpifr}, \ref{aiuc}},    
B.~Klein\inst{\ref{mpifr}},
K.~M.~Menten\inst{\ref{mpifr}}\fnmsep\thanks{Deceased},
O.~Ricken\inst{\ref{mpifr}},
N.~Reyes\inst{\ref{mpifr}},
S.~Sánchez\inst{\ref{iram-es}},
I.~Ruiz\inst{\ref{iram-es}},
C.~Durán\inst{\ref{iram-es}},
D.~John\inst{\ref{iram-es}},
J.~L.~Santaren\inst{\ref{iram-es}},
M.~Sánchez-Portal\inst{\ref{iram-es}},
M.~Bremer\inst{\ref{iram-fr}},
C.~Kramer\inst{\ref{iram-fr}},   %
K.~F.~Schuster\inst{\ref{iram-fr}},
M.~J.~Rioja\inst{\ref{icrar}, \ref{ign}},
R.~Dodson\inst{\ref{icrar}}
        }

   \institute{    
   {Max-Planck-Institut für Radioastronomie, Auf dem Hügel 69, D-53121 Bonn, Germany; \email{gyzhao@mpifr-bonn.mpg.de}}\label{mpifr}
   \and
   {Astronomy Department, Universidad de Concepción, Casilla 160-C, Concepción, Chile}\label{concep}
   \and
   {Institut de Radioastronomie Millimétrique (IRAM), Avenida Divina Pastora 7, Local 20, E-18012, Granada, Spain}\label{iram-es}
   \and
   {Department of Space, Earth and Environment, Chalmers University of Technology, Onsala Space Observatory, SE-43992 Onsala, Sweden}\label{oso}
   \and
   {Finnish Centre for Astronomy with ESO, FI-20014 University of Turku, Finland}\label{turku}
   \and
   {Aalto University Metsähovi Radio Observatory, Metsähovintie 114, FI-02540 Kylmälä, Finland}\label{aalto}
   \and
   {European Southern Observatory, Alonso de Córdova 3107, Vitacura Casilla 7630355, Santiago, Chile}\label{eso-cl}
   \and
   {Centro de Astro-Ingenier{\'i}a, Pontificia Universidad Cat{\'o}lica de Chile, Casilla 306, Santiago, Chile} \label{aiuc}
   \and
   {Institut de Radioastronomie Millimétrique (IRAM), 300 rue de la Piscine, F-38406 Saint Martin d’Hères, France}\label{iram-fr}
   \and
   {International Centre for Radio Astronomy Research, M468, The University of Western Australia, 35 Stirling Hwy, Perth, WA 6009, Australia}\label{icrar}
   \and
   {Observatorio Astron{\'o}mico Nacional (IGN), Alfonso XII, 3 y 5, 28014 Madrid, Spain}\label{ign}
}

\titlerunning{FPT at 86/258\,GHz for APEX--IRAM 30\,m baseline demonstrated}
\authorrunning{G.-Y. Zhao et al.}

   \date{\today}


  \abstract
   {The receiver N3AR operating at a frequency range between 67 and 116\,GHz has been commissioned at the APEX telescope in October 2024.  This adds a new low-frequency band for APEX, with the capability of simultaneous dual-frequency observations using a dichroic beamsplitter.  The 3 mm receiver also allows APEX to join the existing 3 mm global VLBI network.}   
   {One of our commissioning goals was to perform simultaneous dual-band VLBI observations at 86 and 258\,GHz using receivers with shared-optical-paths (SOP) to correct the atmospheric phase fluctuations using the frequency phase transfer (FPT) technique. This was possible together with the IRAM 30 m telescope, which has already developed such a capability.  We aimed to verify the expected phase coherence and sensitivity improvement at the higher frequency achievable by applying FPT.}
   {With the dual-band, single baseline data, we applied the 
   FPT method, which uses the lower frequency data to correct the simultaneously observed higher-frequency data. 
   We evaluate the improvement compared to the conventional single-band observing mode by analyzing the coherence factor in the higher frequency data.}
   {Our results show that the phase fluctuations at the two bands are well correlated. After applying FPT, the interferometric phases at the higher frequency vary much slower and the coherence factor is significantly improved.}
   {Our analysis confirms the feasibility of applying FPT to frequencies above 250\,GHz with SOP receivers. Future observations in this mode could dramatically improve the sensitivity and imaging fidelity of high-frequency VLBI.}


   \maketitle



\section{Introduction}

  Very long baseline interferometry (VLBI) at short millimeter wavelengths has
  provided the sharpest view of active galaxies, including the shadows of supermassive black holes \citep[e.g.,][]{ehtm87p1,ehtsgrap1} and the
  innermost jet region in active galactic nuclei~\citep[AGNs; e.g.,][]{kim2020,janssen2021,issaoun2022,jorstad2023,paraschos2024,baczko2024,lu2023,ros2024}.  
  However, the conventional single-band observing mode still faces severe challenges
  posed by the comparatively lower instrumental efficiencies and the increasingly more turbulent atmospheric conditions at shorter wavelengths~\citep{ehtm87p2,ros2024}. The latter factor typically limits the time coherence of VLBI measurements to $\lesssim 10\,\lambda/\mathrm{mm}$ seconds, with correspondingly poorer fringe detection sensitivity~\citep[e.g.,][]{lobanov2025}.
  Recent developments of multi-band receivers with shared optical paths (SOP)
  present new opportunities to overcome these challenges~\citep[e.g.,][]{han2013}.
  Thanks to the non-dispersive nature of the tropospheric effects, simultaneous dual-/multi-band
  VLBI observations have enabled the application of the frequency phase transfer (FPT)
  method~\citep[e.g.,][]{rd2011, middelberg2005} to calibrate these effects in high-frequency data.
  As a result, the coherence time in the high-frequency data can be extended
  by orders of magnitude after applying FPT from the lower-frequency data.
  These opportunities have been explored by the Korean VLBI Network (KVN) and
  garnered significant global attention in recent years~\citep[][]{rioja2014,rioja2015,algaba2015,yoon2018,zhao2019}.
  Future observations with upgraded global VLBI networks with SOP receivers will
  bring more than an order of magnitude improvements in the sensitivity and dynamic range of mm-VLBI imaging
  and deliver astrometric measurements with an accuracy of a few microarcseconds.
  The resulting exceptional discovery potential would strongly impact a number of scientific fields \citep[e.g.,][]{dodson2017, bonnmeetingreport}.

  FPT has been successfully validated with the KVN at frequencies up to 130\,GHz and baselines of about 500\,km~\citep[e.g.,][]{rioja2015,algaba2015,yoon2018}.
  While FPT is also expected to play an important role for VLBI observations made at
  higher frequencies and longer baselines \citep[e.g.,][]{dodson2017,rioja2023,issaoun2023,jiang2023},
  only limited success has been achieved so far with its applications under such settings \citep{sohn2018}.
  The IRAM 30-m radio telescope located in Pico Veleta (PV), Spain, is able to support simultaneous dual-band operations using the EMIR receiver~\citep{carter2012}, yet the number of partner stations with compatible SOP receivers remains few.
  As an alternative to SOP receivers, co-located antennas can also pair up to
  achieve simultaneous dual-band observations~\citep{asaki1998}.
  Recently, observations at 86 and 215\,GHz were conducted by \citet{issaoun2025} using the IRAM 30\,m telescope (dual-band) and two stations in Hawaii, the James Clerk Maxwell Telescope (JCMT) and the Submillimeter Array (SMA), which are located in close proximity ($\sim$160\,m apart, observing at one band each). 
  The analysis has shown that the visibility phases at the two frequencies are strongly correlated, and the application of FPT could systematically increase the coherence level at the higher frequency. 
  However, the atmospheric conditions above the two stations in Hawaii are not identical due to the difference in their physical locations.
  A rate-removal step, per frequency and scan, was applied prior to the analysis of the correlations of the phases at the two bands and FPT.
  This could limit the application of FPT with the paired-antenna method for weaker sources or require improved schemes for instrumental calibration.

  The Atacama Pathfinder EXperiment~\citep[APEX,][]{gusten2006} is a 12\,meter telescope located in the Atacama desert in Chile at an altitude of 5104 m.
  In September 2024, a new 3 mm APEX receiver (N3AR) developed at the Max-Planck-Institut f\"ur Radioastronomie (MPIfR) in Bonn was successfully installed. 
  N3AR can operate concurrently with the existing receivers in the 230, 345, 460, or 690\,GHz bands (nFLASH230, SEPIA345, nFLASH460, SEPIA690, see~\autoref{sec:n3ar}).  The antenna participated subsequently in the Global mm-VLBI array (GMVA) session \citep[see][]{ros2024} in mid-October 2024.  
  With this upgrade, we could form an inter-continental baseline between APEX and the IRAM 30\,m telescope that is capable of performing simultaneous 86 and 258\,GHz VLBI observations with SOP receivers.
  A test observation was carried out on November 20, 2024.
  Before the dual-band test, both telescopes have operated successfully in single-band mode at 230, 260, and 345\,GHz with the Event Horizon Telescope (EHT) since 2017~\citep{ehtm87p2, raymond2024}. 
  The IRAM 30\,m telescope also participates regularly in the GMVA that operates at 86\,GHz.    
  In this paper, we report the results of the first dual-band test. 

\section{Observations and data analysis} \label{sec:obs_analysis}


    The 86/258 GHz FPT observations took place on November 20, 2024 from 21:00 UT to 01:00 UT
    on the following day with the APEX and the IRAM 30\,m telescopes.
    The baseline length between the two sites was 8,623\,km (\autoref{sec:array_elevation}).     
    The weather conditions were good at both sites during the observation.
    The 225\,GHz zenith opacity, $\tau_{225}$, was around 0.09 and 0.3 at APEX and IRAM 30\,m, respectively.
    The reference frequencies were set to 86.012\,GHz and 258.036\,GHz, with an integer ratio of three between the frequencies.
    APEX had continuous phase-cal tone injection in both receivers for monitoring the instrumental phase drift between the receivers.
    The recording bandwidth was 1024\,MHz at each band.
    APEX recorded dual circular polarization, and IRAM 30\,m telescope recorded dual linear polarization.   

    A total of 8 sources (3C\,454.3, CTA\,102, PKS\,B2201+171, TXS\,2201+315, BL\,Lac, IVS\,B2215+150, PKS\,B0420-014, 3C\,120) were observed.
    We recorded three scans every 30 minutes with a duration of $\sim$5.5 minutes each.
    Two-minute gaps were inserted between scans to slew the antenna and measure the system temperatures.
    The last 8 minutes of each 30-minute block were dedicated to doing pointing checks with bright planets.
    The first 30 minutes (21:00--21:30 UT) were lost due to a power issue with the digital backend at IRAM.
    APEX lost two scans due to an antenna control error, one in the first 30 minutes and another one at 00:37:30 UT.
    In total, we recorded 20 scans of good data of the 8 targets.
    In this paper, we focus on the data of CTA\,102 and 3C\,454.3.
    Each of the two sources was observed in 3 scans, with elevations ranging from 46$^{\circ}$ to 55$^{\circ}$ at APEX and from 47$^{\circ}$ to 25$^{\circ}$ at IRAM 30\,m (see~\autoref{sec:array_elevation}).
    The two sources are close to each other in the sky ($\sim6.8^{\circ}$ apart),
    and the scans are close in time.


    %
    %

  The correlation of the observed data was performed with the DiFX software correlator~\citep{difx2} installed in the
  MPIfR cluster.  
  The frequency resolution was set to 0.5\,MHz and the time-averaging period to 0.4\,s.
  Fringes at both observing bands were successfully detected.   

   \begin{figure}
        \centering
        \includegraphics[width=0.87\linewidth, trim=0.1cm 0.1cm 0cm 0.1cm, clip=true]{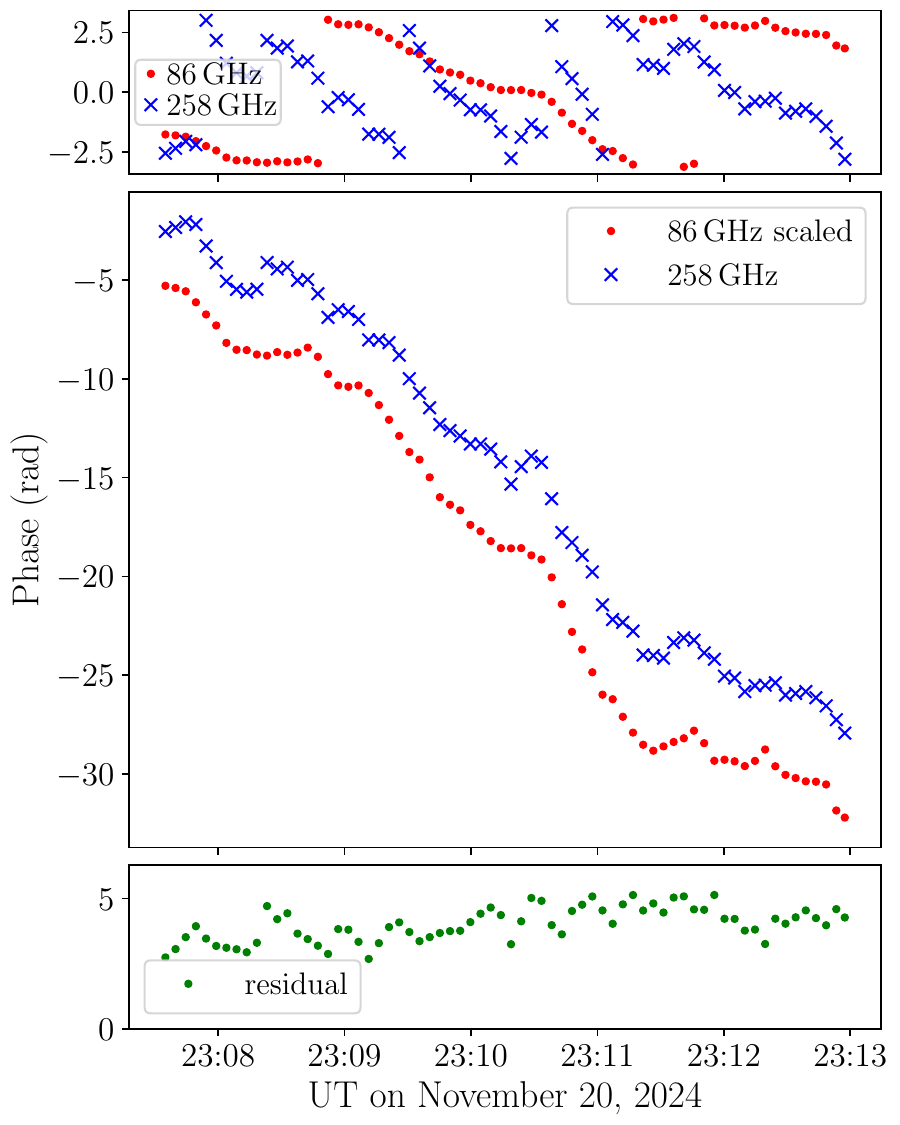}
        \caption{Upper: fringe-fitting phase solutions of the scan at 23:07:30 on CTA\,102 at 86 (red dots) and 258\,GHz (blue crosses); Middle: The unwrapped phase solutions by addressing the 2$\pi$ ambiguities and scaled by the ratio $R$=258\,GHz/$\nu$ for $\nu$=86\,GHz and $\nu$=258\,GHz, respectively; Bottom: The residual between the phases shown in the middle panel. The source elevation was 55$^{\circ}$ and 27$^{\circ}$ at APEX and IRAM 30\,m, respectively.}
            \label{fig:snplt}
    \end{figure}

  The correlated data were analyzed with AIPS in two separate data sets for calibration and the FPT analyses.
  Prior to fringe-fitting, corrections for the Earth orientation parameters and the ionospheric dispersive delays were applied.
  The phase-cal tones for APEX were also loaded and applied to the data.
  We first ran fringe-fitting over a short segment (30 seconds) of the data to initially align the phases across the bands, a step known as manual-pcal.
  We then ran a global fringe-fit over the full data set to determine the multiband delay, delay rate, and phases.
  The solution interval was set to 10 seconds at both bands, and each solution was shifted by 5 seconds from the previous (i.e., SOLINT=10/60; SOLSUB=2). 
  \autoref{fig:snplt} shows the solutions at both bands on one scan of CTA\,102 which was taken after sunset at APEX and around midnight at IRAM. As is visible from the top panel, the higher frequency (258\,GHz) phases vary much faster and could reach 2$\pi$ in about 1 minute, which prevents the detection of faint targets with longer coherent integration times.
  Nevertheless, the phase fluctuations at the higher frequency could be well reproduced by those at the lower frequency (86\,GHz) after being scaled by the frequency ratio (3 in this case, \autoref{fig:snplt}, middle).
  In the FPT step, the phase solutions from the lower frequency band were scaled up by the frequency ratio and applied to the higher frequency.
  The multi-band delay and delay-rate solutions were also applied \citep[see e.g.,][for details of the FPT analysis step]{rioja2015, zhao2019}.


\section{Results} \label{sec:results}
\begin{figure*}
            \centering
            \includegraphics[width=0.87\linewidth, trim=0cm 0.2cm 0cm 0.2cm, clip=true]{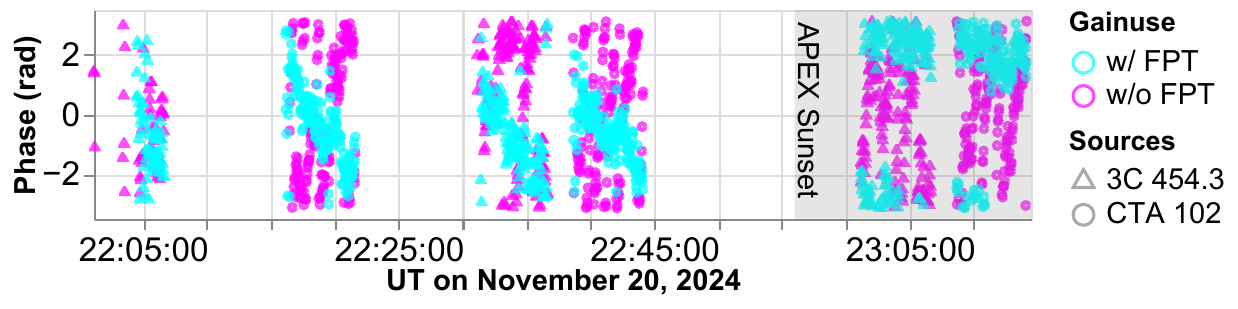}
            \caption{The interferometric phases of CTA\,102 and 3C\,454.3 at 258.036\,GHz before (magenta) and after (cyan) applying FPT. The data are averaged over 2 seconds and across the bandwidth. The shaded area indicates the time after sunset at APEX.} 
                \label{fig:vplot}
        \end{figure*}

  As shown in~\autoref{fig:snplt}, the phases at the two frequencies are well correlated.
  We can expect the visibility phases after applying FPT would show improved data quality.
  In~\autoref{fig:vplot}, the phases of CTA\,102 and 3C\,454.3 at 258\,GHz before and after FPT are shown.
  As is obvious from the figures, the phases before FPT, which are equivalent to the phases with only single-band observations, show much faster phase variations, indicating the phases are corrupted by short-time scale propagation effects.
  However, the strong correlation between bands and the 1:3 ratio of the phase rates in \autoref{fig:snplt} indicate that the dominating propagation effects are non-dispersive and can be corrected by FPT.
  The phases after FPT exhibit only slow variations.
  The range of the phase fluctuations within a scan after sunset reduces from 26\,rad before FPT to 1.2\,rad after FPT, which corresponds to a scan-averaged phase rate of 13 and 0.6\,mHz before and after FPT, respectively. For the scans before sunset, the ranges change from 16$\sim$30\,rad to 3$\sim$5\,rad.
  The residual phases of the two sources on neighboring scans have similar phase rates,
  which hints at the possibility of using one source to calibrate the other, i.e., the so-called source-frequency phase referencing method (see~\autoref{sec:sfpr}).

\begin{figure}
       \centering
       \includegraphics[width=0.85\hsize]{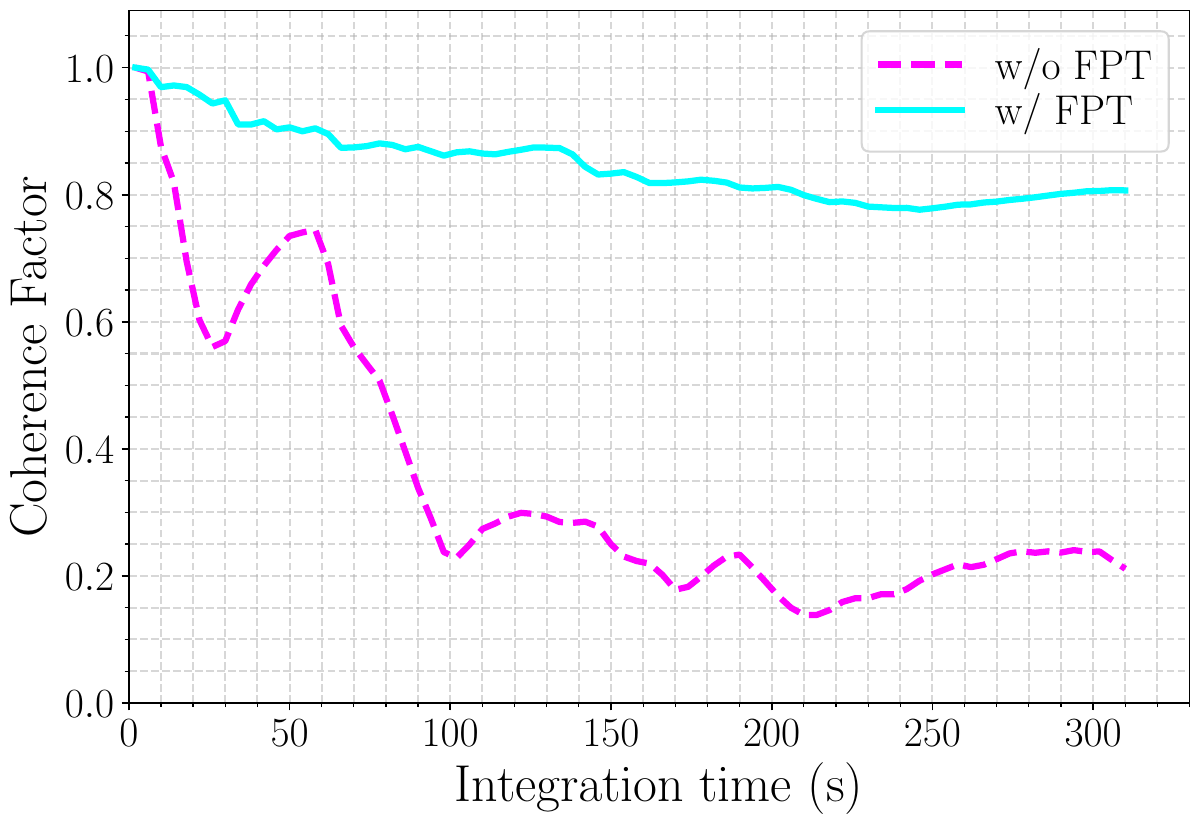}
          \caption{Coherence factor before and after applying FPT for the CTA\,102 scan at 23:07:30.}
             \label{fig:coherence}
       \end{figure}

  To quantify the coherence of the data and compare the cases before and after FPT,
  we analyze the coherence factor 
  \begin{equation}
    C(T)=\left|\frac{1}{T}\int_{0}^{T}\exp[i\phi(t)]dt\right|
    \label{eq:coherence}
  \end{equation}
  where $T$ is the time interval in which the coherence factor is examined,
  and $\phi(t)$ is the visibility phase at time $t$~\citep{TMS2017}.
  The coherence factor is defined to be 1.0 when the data are perfectly coherent.
  In~\autoref{fig:coherence}, the coherence factors before and after FPT are shown
  for the scan of CTA\,102 starting at 23:07:30 UT (as shown in \autoref{fig:snplt}).
  The coherence factor curve without FPT decreases steeply with increasing integration time.
  The value drops below 0.9 when the integration time is longer than $\sim$10 seconds,
  and reaches 0.6 at $\sim$20 seconds.
  On the contrary, the curve resulting from applying FPT shows a milder decline over time.
  At all integration intervals within the scan, the values stay above the curve obtained without FPT (see also Figures~\ref{fig:coherence_cta_extra}--\ref{fig:coherence_0420}).
  For integration time within one minute, the coherence factor stays higher than 0.9
  and remains around 0.8 for longer intervals up to the scan length of 5 minutes.
  Improvement of the coherence by FPT can also be found in other scans and sources (see~\autoref{sec:other_scans}).
  
    %

\section{Discussion} \label{sec:discussion}

  The short timescale tropospheric turbulence has posed major challenges to
  conventional single-band VLBI observations at mm-wavelengths.
  Nonetheless, the non-dispersive nature of these effects has motivated significant
  developments of SOP receivers that could enable the application of FPT to calibrate these effects.
  Successful validations at wavelengths longer than 2\,mm have been achieved in the past decades.
  Simultaneous dual-band observations have now become available at shorter wavelengths
  with facilities capable of conducting VLBI observations at the highest frequencies so far~\citep[345\,GHz, see][]{raymond2024}.
  While the non-dispersive property would still be valid in this frequency range,
  successful applications of FPT would rely on the instrumental phase stability of the participating stations.
  Our present results have thus confirmed that the current state-of-the-art design of SOP receivers is valid at high frequencies.

  The dispersive propagation phase terms, like ionospheric and instrumental phases,
  along with the source structure, are the main sources of residual phases after FPT.
  We note that on several scans of our observations, the FPT residual phases vary
  faster than those seen at lower frequency pairs~\citep[e.g.,][]{rioja2015, algaba2015},
  whereas the expected ionospheric effects are much less at higher frequencies.
  The observations occurred during nighttime at the IRAM 30\,m site. However, the difference before and after sunset at APEX (see~\autoref{fig:vplot}) still indicates noticeable ionospheric contributions during daytime. 
  This can be explained as the solar activity was high in 2024~\citep[e.g.,][]{jouve2025}, and the APEX site is near the geomagnetic equator.
  The similarity of the phases between sources suggests it is less likely to be dominated by source structure.
  One potential instrumental cause would be the alignment of the primary beams at the two frequencies
  at APEX because the two receivers are located in different cabins and focus positions.
  Another potential cause is phase drift between the receivers. APEX had continuous tone injection to measure and correct the inter-receiver phase offset, but IRAM 30\,m lacked such a system that could operate during the astronomical observation.
  Compared to the recent work by Issaoun et al. (2025), the coherence factor without FPT in our case is lower (reaching $\sim$0.2 compared to $\sim$0.6). This is mainly because we do not need to apply the rate-removal process, and the sky opacity is higher in our case ($\tau_{225}\sim$0.3 compared to 0.25 at IRAM 30\,m). However, the coherence after FPT is comparable in the two cases, despite the higher frequency and sky opacity in our case. 
  This could also point to common instrumental contributions as the limiting factors and underscores the need for continuous tone injection in both bands to stabilize the relative phase between the receivers.
  We note that the beam alignment at APEX is expected to be improved in March 2025, and tone injection at IRAM 30\,m is being reconfigured to permit continuous injection during observations,
  so future observations might show reduced residual phases and further improved phase coherence.

  Recent works have demonstrated the feasibility of applying FPT on intercontinental baselines and at high frequencies~\citep[e.g.,][and this work]{zhao2019, issaoun2025}.
  In the future, the use of FPT will become more common for mm-VLBI observations with global VLBI networks.
  FPT mode will become a regular part of the GMVA sessions in the coming years~\citep{lobanov2025}.  
  At frequencies above 150\,GHz, the increasing number of EHT telescopes with SOP receivers will greatly improve the baseline sensitivity and image dynamic ranges \citep[an order of magnitude higher, see e.g., ][]{lobanov2025, bonnmeetingreport} of future observations.
  Furthermore, the APEX receiving system could also support simultaneous observations at 86 and 345\,GHz.       
  An FPT test could be scheduled in the near future with, e.g., the paired antennas in Hawai'i or another upcoming station with SOP receivers.   



\section{Conclusions and future aspects}

  Motivated by the installation of the new N3AR receiver at APEX and the existing SOP receivers at the IRAM 30\,m telescope~\citep{carter2012},
  we have performed the first simultaneous 86 and 258\,GHz VLBI observations.
  The observations were carried out on November 20, 2024, and dual-band fringes were successfully obtained.

  Our analyses have shown that FPT is successfully applied up to 258 GHz, which is the highest frequency at which this method has been performed on an inter-continental baseline.
  The coherence time is extended from $\sim$10 seconds to $\sim$1 minute by FPT.
  The residual phases between nearby sources also trace each other.
  This success paves the way for more FPT tests in the future with sufficiently many telescopes, enabling high dynamic range imaging and high precision (micro-arcsecond level) relative astrometry at high frequencies.

\begin{acknowledgements}
We are grateful to all the scientific and technical staff involved in the commissioning of the APEX N3AR receiver and in previous dual-band tests with IRAM. We also thank Dr. Sergio A. Dzib-Quijano for carefully reading the manuscript and providing helpful comments. 
 This work was supported by the European Union Horizon 2020 research and innovation programme under grant agreements RadioNet (No. 730562), M2FINDERS (No. 101018682).
APEX is a collaboration between the Max-Planck-Institut f\"ur Radioastronomie (Germany), ESO, and the Onsala Space Observatory (Sweden).
The IRAM 30-m telescope on Pico Veleta, Spain is operated by IRAM and supported by CNRS (Centre National de la Recherche Scientifique, France), MPG (Max-Planck-Gesellschaft, Germany), and IGN (Instituto Geográfico Nacional, Spain).
The quarter-wave plate used at APEX was designed by Anne-Laure Fontana at IRAM.  

This research has made use of the following software packages:
DiFX~\citep{difx2}, AIPS~\citep{aips}, ParselTongue~\citep{parseltongue},
SMILI~\citep{smili}, Matplotlib~\citep{matplotlib}, Altair~\citep{altair},
pandas~\citep{pandas}, numpy~\citep{numpy}, GeoPandas~\citep{geopandas}.
\end{acknowledgements}

%

\bibliographystyle{aa}
\bibliography{apex_pv.bib}

\begin{appendix}
  \onecolumn
  \section{N3AR - a new 3\,mm receiver and dual-band capability at the APEX telescope} \label{sec:n3ar}
       
       \begin{figure}
            \centering
            \includegraphics[width=0.92\linewidth, trim=0.05cm 0.0cm 0.0cm 0.03cm, clip=true]{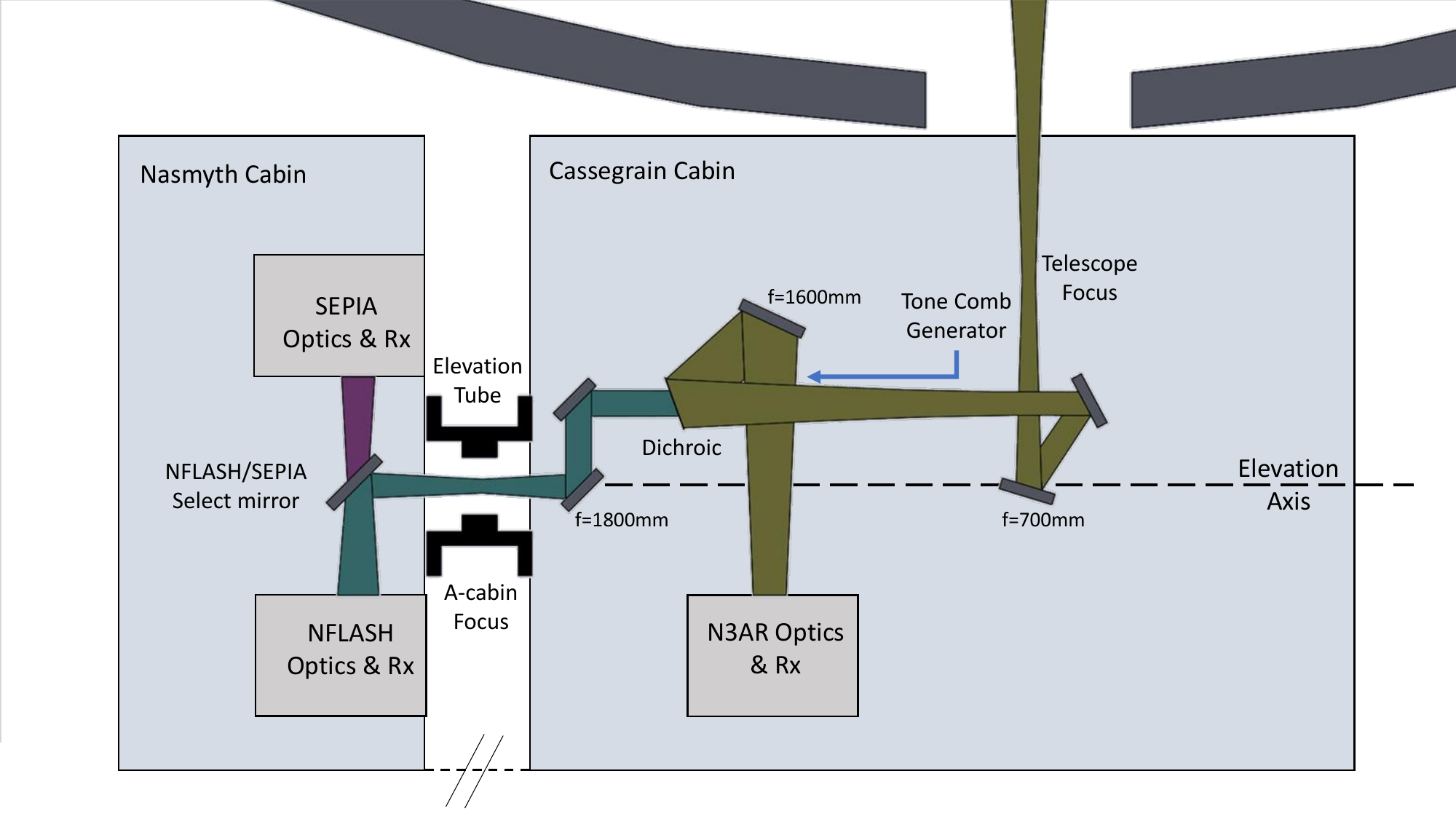}
            \caption{Optical path and block schematic of the APEX receivers.}
            \label{fig:n3ar}
        \end{figure}
 

    The new 3\,mm APEX receiver (N3AR) was designed and built by the Sub-mm Group at the MPIfR.  It was installed
    at the APEX telescope in September 2024 and commissioned in October 2024 with VLBI observations in the GMVA
    at 86\,GHz.  The dual-band capability was commissioned in November 2024 with the shared-optics FPT
    observation described in this paper.  
    
    The receiver optical path and block schematic are shown in \autoref{fig:n3ar}.  The relay
    optics are designed to illuminate the secondary mirror with an edge taper of 12.5\,dB.  The beam clearance at the edge of the optical elements is 4 omega.

    The receiver has cooled high-electron-mobility transistor (HEMT) amplifiers covering the band 67 to 116\,GHz.  The Dewar output is at radio frequency (RF), and downconversion to an intermediate frequency (IF) of 1.5--20\,GHz is performed in a room--temperature sideband--separating downconverter module from RPG (Radiometer Physics GmbH).
    The local oscillator (LO) is derived from a VDI synthesizer in the range 8--20\,GHz followed by a $\times$8 multiplier chain.  
    The receiver provides a total of 4 $\times$ 18\,GHz IF bands, being dual native-linear polarizations and two sidebands.
    %
    Calibration of receiver temperature ($T_{rec}$) is performed with hot (348\,K) and cold (290\,K) loads. Switching optics can send the beam to the sky or hot or cold load, where the hot load is a heated, thermally isolated, absorber-coated cone, and the cold load is an absorber made of TK Instrument's RAM tiles at cabin ambient temperature.
    
    A dichroic beamsplitter is inserted in the existing optical path in the Cassegrain cabin to reflect the 3\,mm band to the N3AR receiver and allow higher frequencies to pass to the existing Nasmyth A Cabin receivers nFLASH-230 or SEPIA-345~\citep{belitsky2018, meledin2022}.  
    Those cover the bands 196--281\,GHz and 272--376\,GHz respectively, any of which can be operated simultaneously with N3AR.
    The dichroic beamsplitter consists of two parallel, cross-wired grids.  The wire pitch (500\,{\textmu}m) sets the 3\,mm reflectivity, and the separation between the two grids sets the frequency at which peak transmission occurs.  
    The N3AR receiver is optically aligned with the nFLASH receiver to achieve concentric beams on the sky.  
    
\begin{figure}
       \centering
       \includegraphics[width=0.95\hsize]{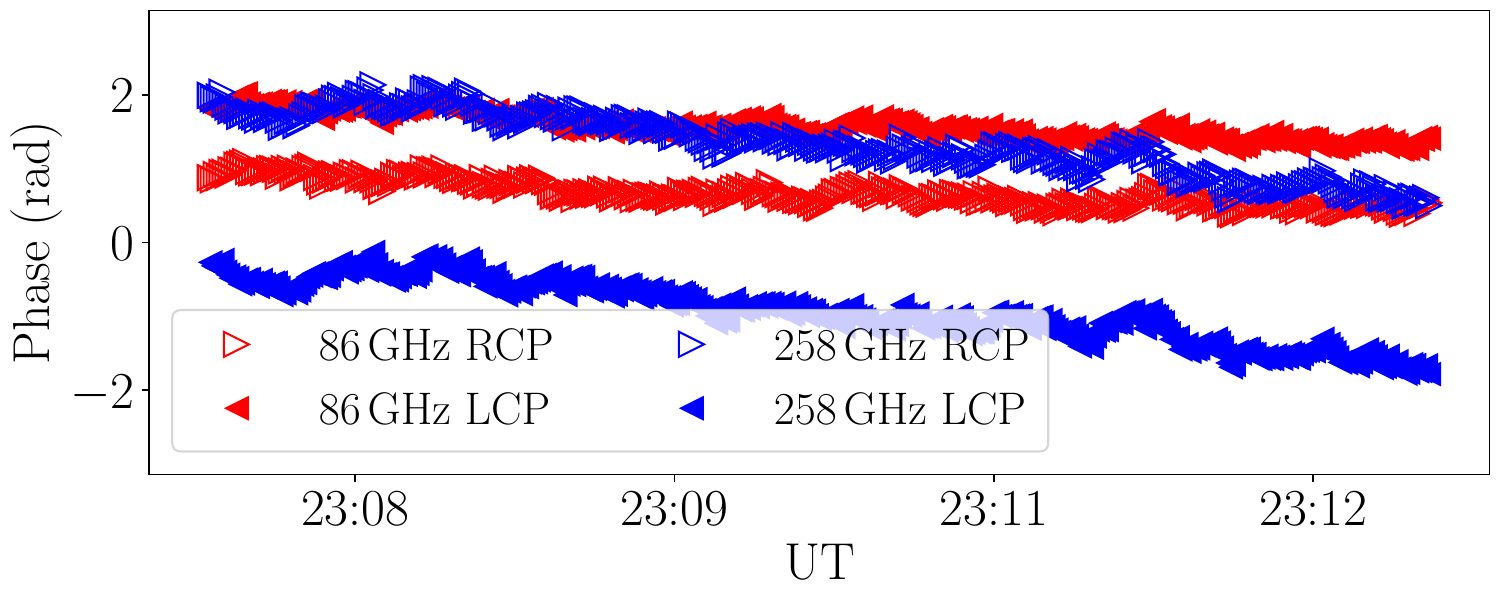}
          \caption{APEX phase-cal tones phases at the two observed frequencies the CTA\,102 scan at 23:07:30.}
             \label{fig:pcal-2307}
       \end{figure}

    A frequency comb for receiver phase stabilization is generated with a Schottky diode mixer  (Pacific Millimeter WM) pumped in the range 14 to 19\,GHz with an Agilent 8257D synthesizer and amplifier Microsemi AML618P2501 and radiated with a standard pyramidal horn alongside the beam from approximately the 5 omega border of the astronomical beam.
    The tone power drops off strongly with frequency. 
    This required an equalizer, realized as a waveguide filter, to attenuate the 3\,mm band to prevent saturation of N3AR while having enough power in nFLASH at 258\,GHz.
    The comb is needed to remove the relative instrumental phase drift between N3AR and the Nasmyth cabin receivers.  
    The drift occurs due to different temperature environments and different cable runs transporting the maser 10\,MHz reference.  
    The phase-cal tone is recorded along with the astronomical signal with the VLBI backend and the tone phase time series is extracted in the correlator for both receiver bands and used to correct the astronomical visibility phase time series.  
    An example of the tone phase time series is shown in \autoref{fig:pcal-2307}, and displays phase rates of $\sim$-5 and -14\,deg/min at 86 and 258\,GHz.
    
    A quarter-wave plate provides conversion between dual circular polarization on the sky for VLBI and dual linear polarization that is native for the receiver.  The plate was designed at IRAM in grooved Teflon and was used for the Plateau de Bure interferometer prior to upgrade as NOEMA.  
    
    \twocolumn



  \section{Antenna locations and source elevations} \label{sec:array_elevation}
\begin{figure}
       \centering
       \includegraphics[width=0.95\hsize]{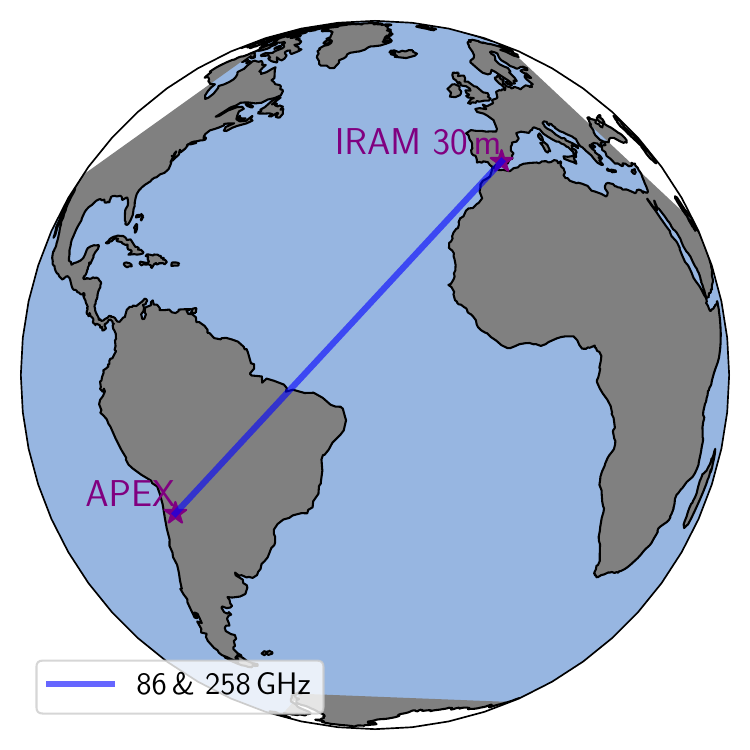}
          \caption{The location of the antennas participated in the 86 \& 258\,GHz dual-band VLBI observations on November 20, 2024.}
             \label{fig:baseline}
       \end{figure}

  In \autoref{sec:obs_analysis}, we described the first simultaneous 86 \& 258\,GHz dual-band observations on November 20, 2024. The observations were jointly carried out with the APEX and IRAM 30-m telescope. 
  In \autoref{fig:baseline}, we show the locations of the telescopes on a map.

  Eight sources with right ascension between 22h00m and 04h00m were observed.
  All sources were detected previously with the GMVA at 86\,GHz~\citep{nair2019}.
  Five of the targets (3C\,454.3, CTA\,102, BL\,Lac, PKS\,B0420-014, 3C\,120) have a total flux higher than 1.5\,Jy~\citep[e.g,][]{nair2019}. The other three (PKS\,B2201+171, TXS\,2201+315, IVS\,B2215+150) are fainter but are close to the bright sources ($< 11^{\circ}$).
  The 3C\,454.3 and CTA\,102 results shown in this paper were obtained during UT 22:00 -- 23:15. A summary of the target source, UT range, and elevations of each scan can be found in \autoref{tab:scan_summary}.  

    %

    \begin{table*}
    \centering
    \caption{Scan summary of the 86/258\,GHz FPT test on November 20, 2024}
    \label{tab:scan_summary}
    \begin{tabular}{ccccccc}
        \hline
        \textbf{Scan} & \textbf{Source} & \textbf{Start} & \textbf{End} & \textbf{APEX} & \textbf{IRAM 30m} & \textbf{Notes} \\
        \textbf{No.} & &  &  & \textbf{elevation} & \textbf{elevation} & \\
        \textit  &  & (UT) & (UT) & (deg) & (deg) &  \\
        \hline 
        1 & 3C\,454.3 & 21:00:00 & 21:05:30 & 37 -- 38 & 58 -- 57 & IRAM digital backend power issue \\
        2 & CTA\,102 & 21:07:30 & 21:12:30 & 45 -- 46 & 50 -- 49 &   ...; APEX control error\\
        3 & PKS\,B2201+171 & 21:15:00 & 21:20:30 & 45 -- 46 & 46 -- 45 &  IRAM digital backend power issue \\
        \hline
        4 & BL\,Lac & 21:30:00 & 21:35:00 & 23 -- 23 & 53 -- 52 &  \\
        5 & TXS\,2201+315 & 21:37:30 & 21:42:30 & 33 -- 34 & 49 -- 48 &  \\
        6 & BL\,Lac & 21:45:00 & 21:50:00 & 24 -- 24 & 50 -- 49 &  \\
        \hline
        7 & 3C\,454.3 & 22:00:00 & 22:05:30 & 46 -- 46 & 47 -- 46 &  \\
        8 & IVS\,B2215+150 & 22:07:30 & 22:13:00 & 50 -- 51 & 38 -- 37 &  \\
        9 & CTA\,102 & 22:15:00 & 22:20:30 & 53 -- 54 & 37 -- 36 &  \\
        \hline
        10 & 3C\,454.3 & 22:30:00 & 22:35:30 & 49 -- 49 & 41 -- 40 &  \\
        11 & CTA\,102 & 22:37:30 & 22:43:00 & 55 -- 55 & 33 -- 31 &  \\
        12 & IVS\,B2215+150 & 22:45:00 & 22:50:30 & 52 -- 52 & 30 -- 29 &  \\
        \hline
        13 & 3C\,454.3 & 23:00:00 & 23:05:30 & 50 -- 50 & 35 -- 34 &  \\
        14 & CTA\,102 & 23:07:30 & 23:13:00 & 55 -- 55 & 27 -- 26 &  \\
        15 & PKS\,B2201+171 & 23:15:00 & 23:20:30 & 48 -- 48 & 23 -- 21 &  \\
        \hline
        16 & BL\,Lac & 23:30:00 & 23:35:00 & 23 -- 23 & 32 -- 31 &  \\
        17 & TXS\,2201+315 & 23:37:30 & 23:42:30 & 33 -- 33 & 26 -- 25 & APEX control error \\
        18 & BL\,Lac & 23:45:00 & 23:50:00 & 23 -- 22 & 29 -- 28 &  \\
        \hline
        19 & BL\,Lac & 00:00:00 & 00:05:00 & 22 -- 21 & 26 -- 26 &  \\
        20 & TXS\,2201+315 & 00:07:30 & 00:12:30 & 31 -- 30 & 20 -- 19 \\
        21 & BL\,Lac& 00:15:00 & 00:19:00 & 20 -- 20 & 24 -- 23 &  \\
        \hline
        22 & PKS\,B0420-014 & 00:30:00 & 00:35:00 & 23 -- 24 & 52 -- 52 &  \\
        23 & 3C\,120 & 00:37:30 & 00:42:30 & 19 -- 20 & 58 -- 58 &  \\
        24 & PKS\,B0420-014 & 00:45:00 & 00:50:00 & 26 -- 27 & 52 -- 52 &  \\
        \hline
      \end{tabular}
     \end{table*}

  
  \section{Source-frequency phase referencing} \label{sec:sfpr}

    FPT could calibrate the rapidly varying tropospheric phase errors in the higher frequency band, along with all other non-dispersive phase effects, like those due to antenna and source position errors.
    However, the dispersive propagation effects, like ionospheric and instrumental phase errors, will remain in the FPT residual data.
    Nevertheless, these effects vary slowly in time -- meaning that FPT will still extend the coherence time (see \autoref{sec:results}) -- making it possible to perform fringe-fitting with a much longer integration time.
    Alternatively, it is also possible to calibrate these remaining effects by phase referencing the FPT residual phase to that of a nearby source because these effects also change only mildly with direction.
    This method is called source-frequency phase referencing (SFPR) and would require alternating the scans between two or more sources.
    SFPR can not only calibrate all the remaining propagation effects, but it also preserves the astrometric information.
    For details, please refer to e.g., ~\citet{rd2011, rd2020}.

    %

\begin{figure}
            \centering
            \includegraphics[width=\linewidth]{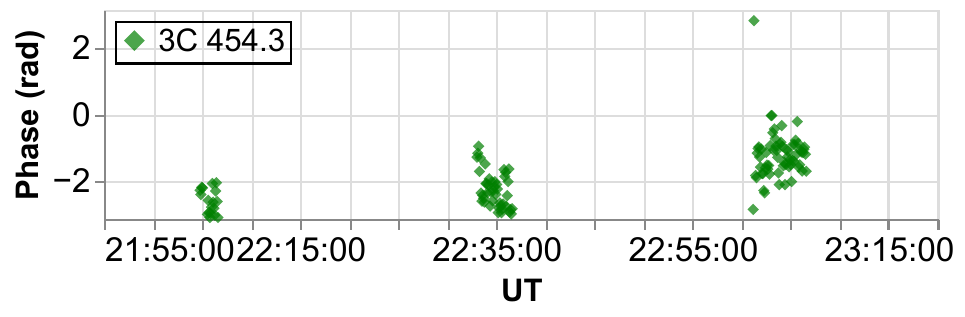}
            \caption{The interferometric phases of 3C\,454.3 at 258.036\,GHz after applying SFPR. The reference source for SFPR was CTA\,102. The FPT was performed from the 86.012\,GHz data of the same source. The data are averaged over 5 seconds.}
            \label{fig:sfpr}
        \end{figure}

    In our experiment, the targets CTA\,102 and 3C\,454.3 were $\sim$6.8 degrees from each other.
    Our results confirmed that the FPT residual phases of the two sources follow a similar trend.
    The average phase rates at 258\,GHz data after FPT for both sources are -1.2$\sim$-1.7\,mHz and $\sim$-0.6\,mHz before and after APEX sunset. 
    In this appendix, we show the results from calibrating the dispersive effects in the FPT residual of 3C\,454.3 using CTA\,102 as the reference source.
    The phases after this step are shown in \autoref{fig:sfpr}.
    The residual phases after SFPR show a smooth variation of 1.5\,radian in 1 hour, which corresponds to a phase rate of 0.07\,mHz.
    The SFPR residual phases should consist only of the information on source structure, source position, and interpolation error~\citep[see e.g.,][]{rd2011}.
    However, we want to note that in our experiment, the effective switching cycle between sources is 30 minutes.
    This time is much longer than the coherence time after FPT, so the calibration would be suboptimal.
    Based on our findings that the coherence factor is around 0.8 at 5 minutes,
    and considering the actual slewing speed of the antennas and the overhead time required to measure system temperatures and so on would require $\sim$1.5 minutes,
    we would recommend that a source switching cycle of 5 to 10 minutes for future SFPR tests in this frequency range.
    Future tests with more baselines would also provide constraints on the source structure using closure quantities~\citep[e.g.,][]{blackburn2020}.
    Altogether, this will result in high-precision astrometric measurements of the ultracompact sources at the highest frequencies~\citep[e.g.,][]{rd2020,jiang2023,lobanov2025}.

    \FloatBarrier

  \section{FPT and coherence analysis on other scans and sources}~\label{sec:other_scans}
\begin{figure}
            \centering
            \includegraphics[width=\linewidth]{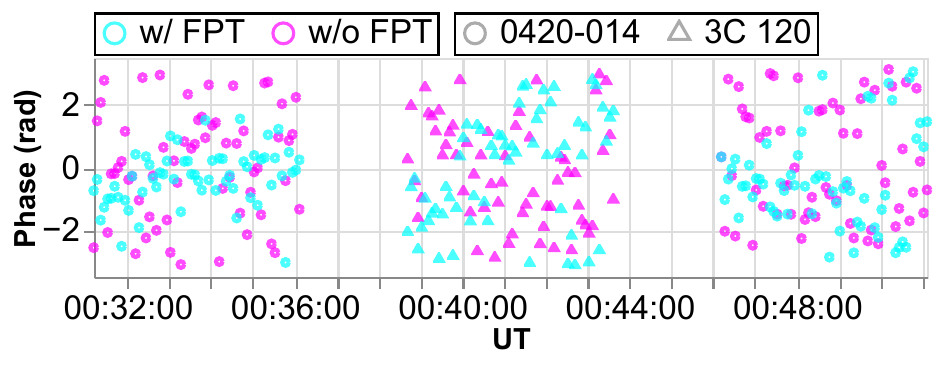}
            \caption{Same as~\autoref{fig:vplot} but for PKS\,B0420-014 and 3C\,120. The data were averaged in 5\,seconds. The longer averaging time compared to that for CTA\,102 and 3C\,454.3 is for reducing the thermal noise in the data.}
            \label{fig:vplot_0420_3c120}
        \end{figure}

    Besides CTA\,102 and 3C\,454.3, six more sources were also observed in our FPT test (see~\autoref{sec:array_elevation}). 
    We briefly summarize the data properties here. 
      \begin{itemize}
          \item PKS\,B0420-014 and 3C\,120 were observed in the last 30 minutes (00:30--01:00\,UT). For these two sources, fringes at 258\,GHz could not be directly detected with a sufficient signal-to-noise ratio (s/n). Nevertheless, the data could be recovered after applying FPT from 86\,GHz. In~\autoref{fig:vplot_0420_3c120}, we show the interferometric phases of these two sources at 258\,GHz before and after applying FPT. Same as in~\autoref{fig:vplot}, FPT from 86\,GHz could significantly improve the coherence in the high frequency data.
          The phase rates after FPT were 0.5, 2.4, and -1.9\,mHz, for the three scans respectively, while the data before FPT are dominated by noise.
          We note, however, in the case of 3C\,120, the data are still affected by comparatively large thermal noise due to the low elevation of the source at APEX during the scan (19$^{\circ}$).
          
          \item For the three faint targets, PKS B2201+171, TXS 2201+315, and IVS B2215+150, few fringe detections at 86\,GHz were obtained with the same setups (i.e., 10-second solution interval, s/n threshold of 5), which prevents us from applying FPT.
          
          \item BL\,Lac was observed on 6 scans during 21:30--22:00UT and 23:30--00:30 UT. While we also see significant coherence improvements at 258\,GHz after applying FPT, the residual phases still show faster variations, roughly three times that of the other four sources (CTA\,102, 3C\,454.3, PKS\,B0420-014, \& 3C\,120).
          This fast variation might have a source-intrinsic origin, indicating an offset between the emission region at the two frequencies, or an ionospheric and/or instrumental origin as discussed in~\autoref{sec:discussion} due to the low elevation. Further analysis is necessary to better understand the data, and the results will be presented in forthcoming publications.
          
      \end{itemize}

\begin{figure}
       \centering
       \includegraphics[width=0.90\hsize]{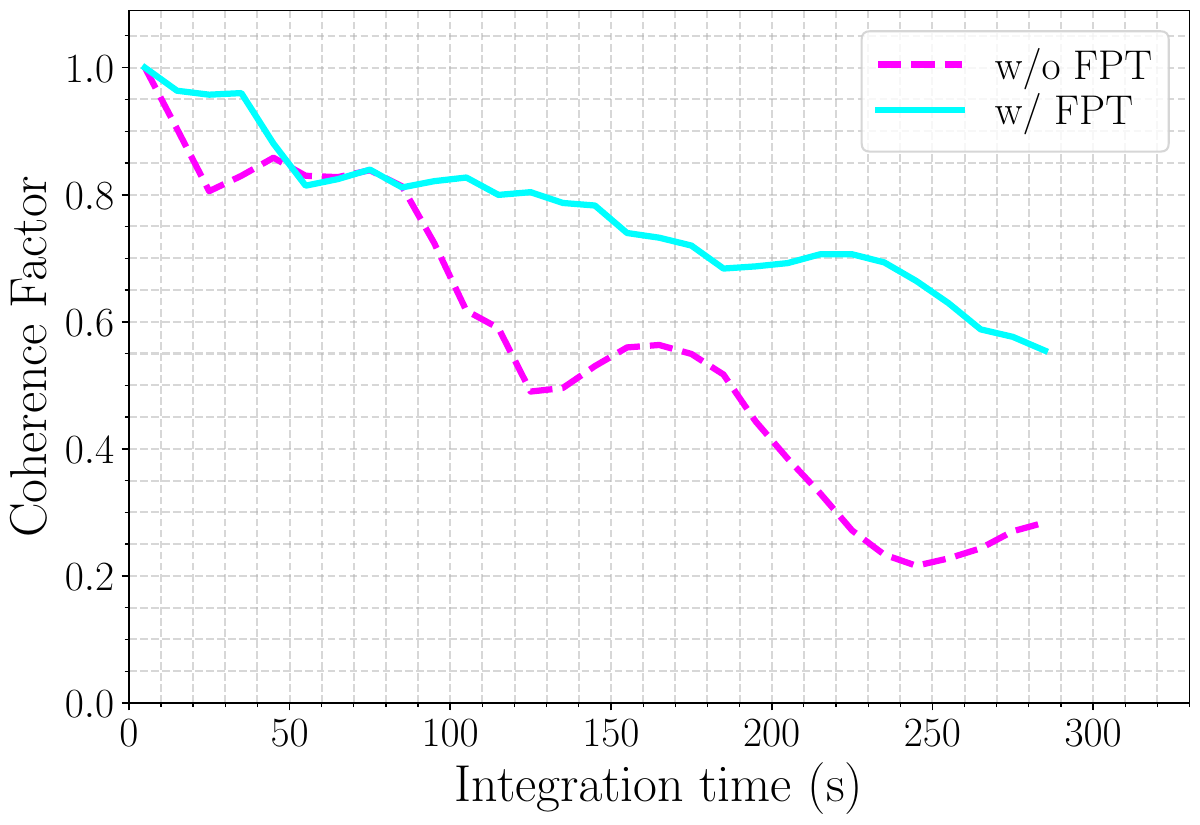}
       \includegraphics[width=0.90\hsize]{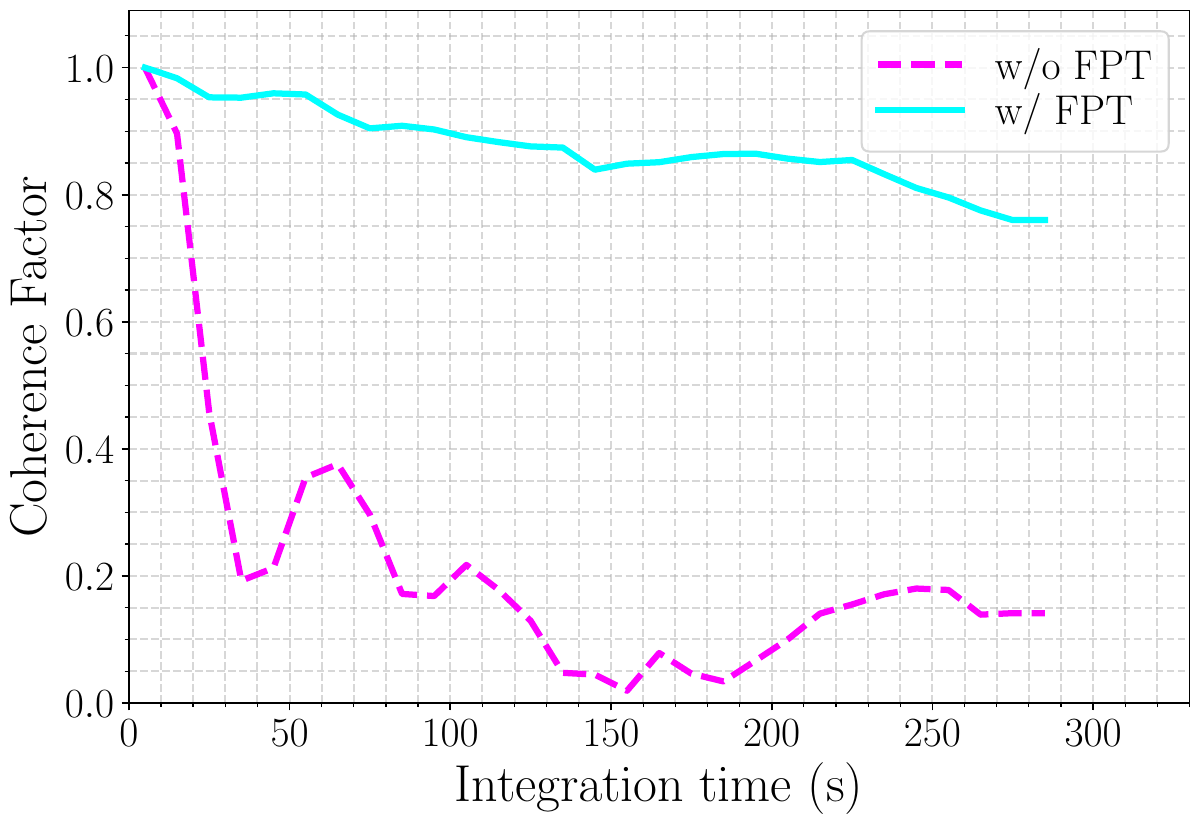}
          \caption{Same as \autoref{fig:coherence} but for the other CTA\,102 scans at 22:15:00 (top) and 22:37:30 (bottom), respectively.}
             \label{fig:coherence_cta_extra}
       \end{figure}

\begin{figure}
       \centering
       \includegraphics[width=0.90\hsize]{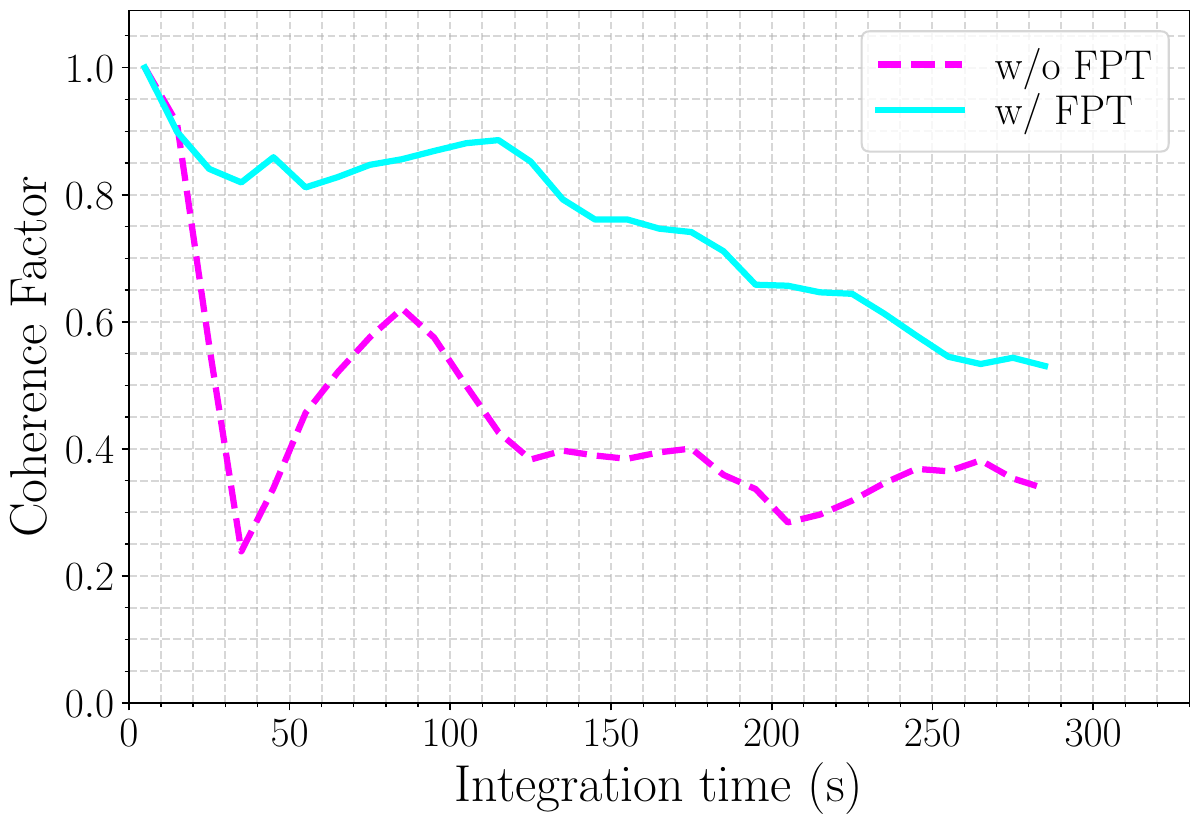}
       \includegraphics[width=0.90\hsize]{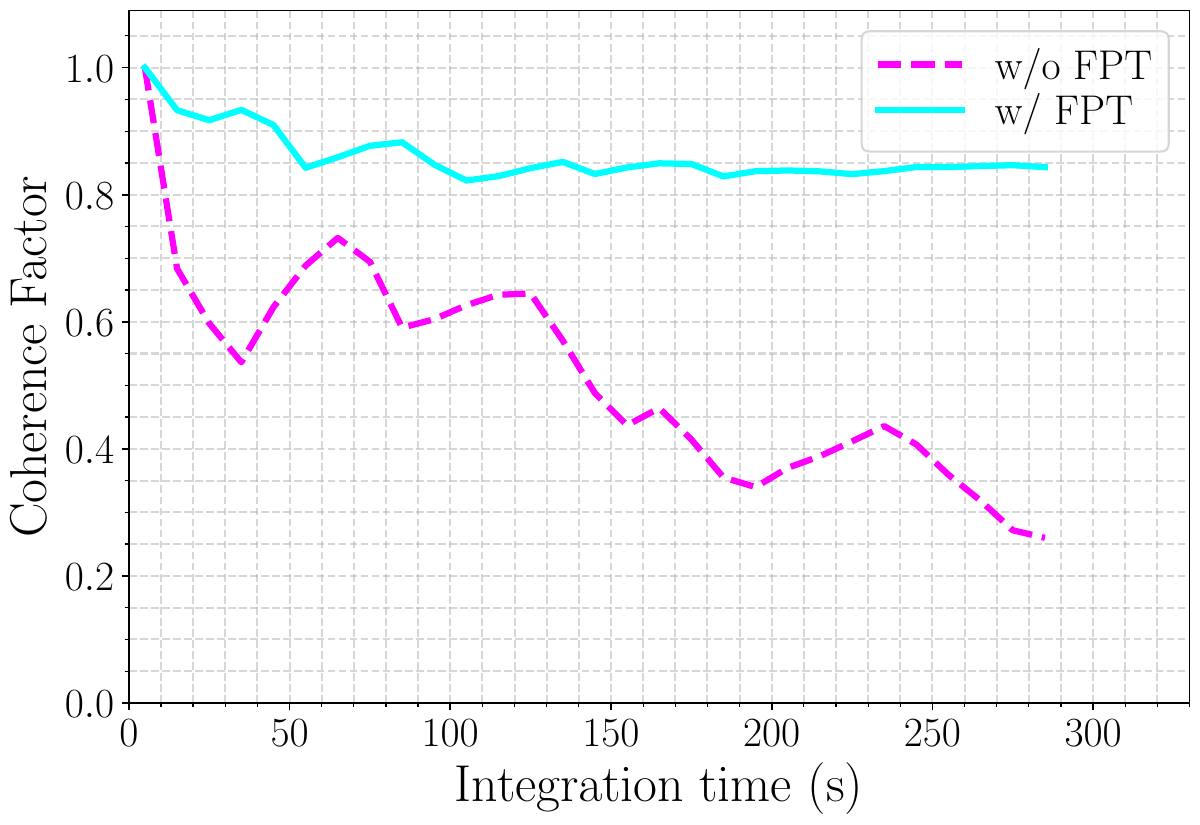}
          \caption{Same as \autoref{fig:coherence} but for the 3C\,454.3 scans at 22:30:00 (top) and 23:00:00 (bottom), respectively.}
             \label{fig:coherence_3c454}
       \end{figure}

\begin{figure}
       \centering
       \includegraphics[width=0.90\hsize]{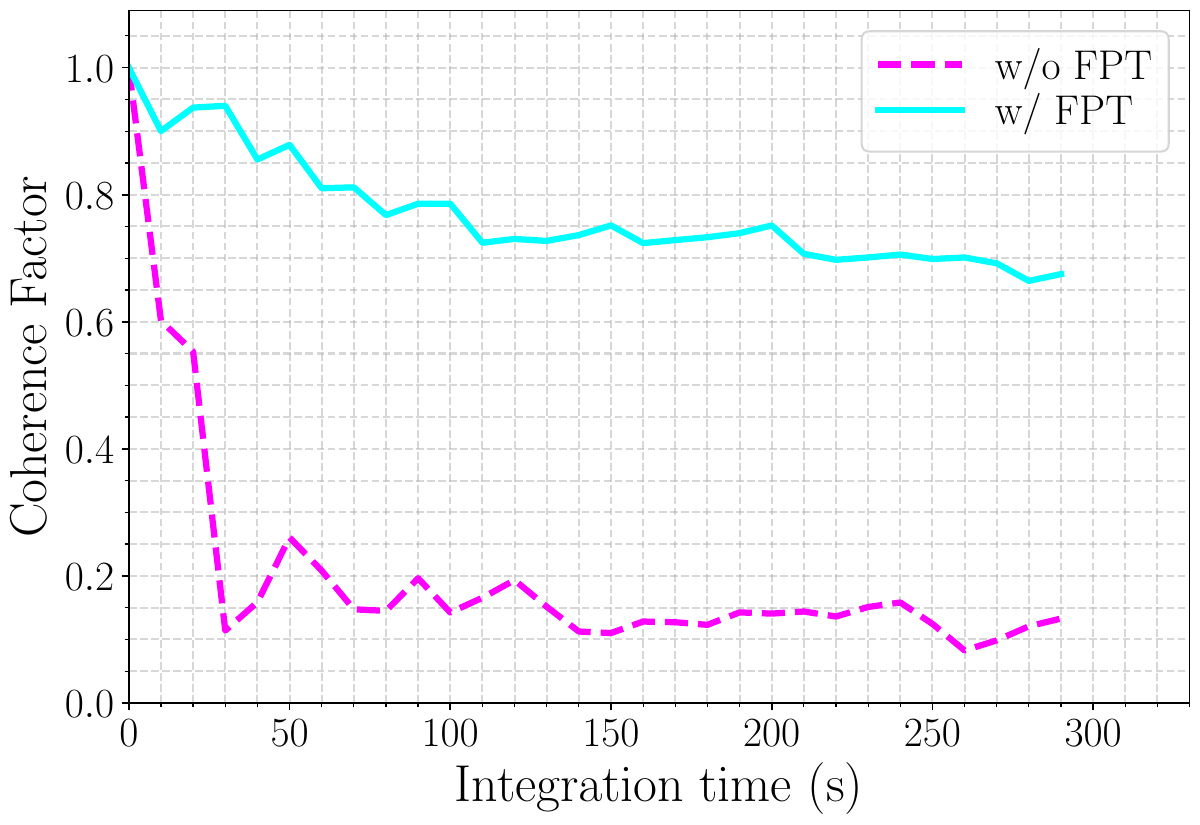}
       \includegraphics[width=0.90\hsize]{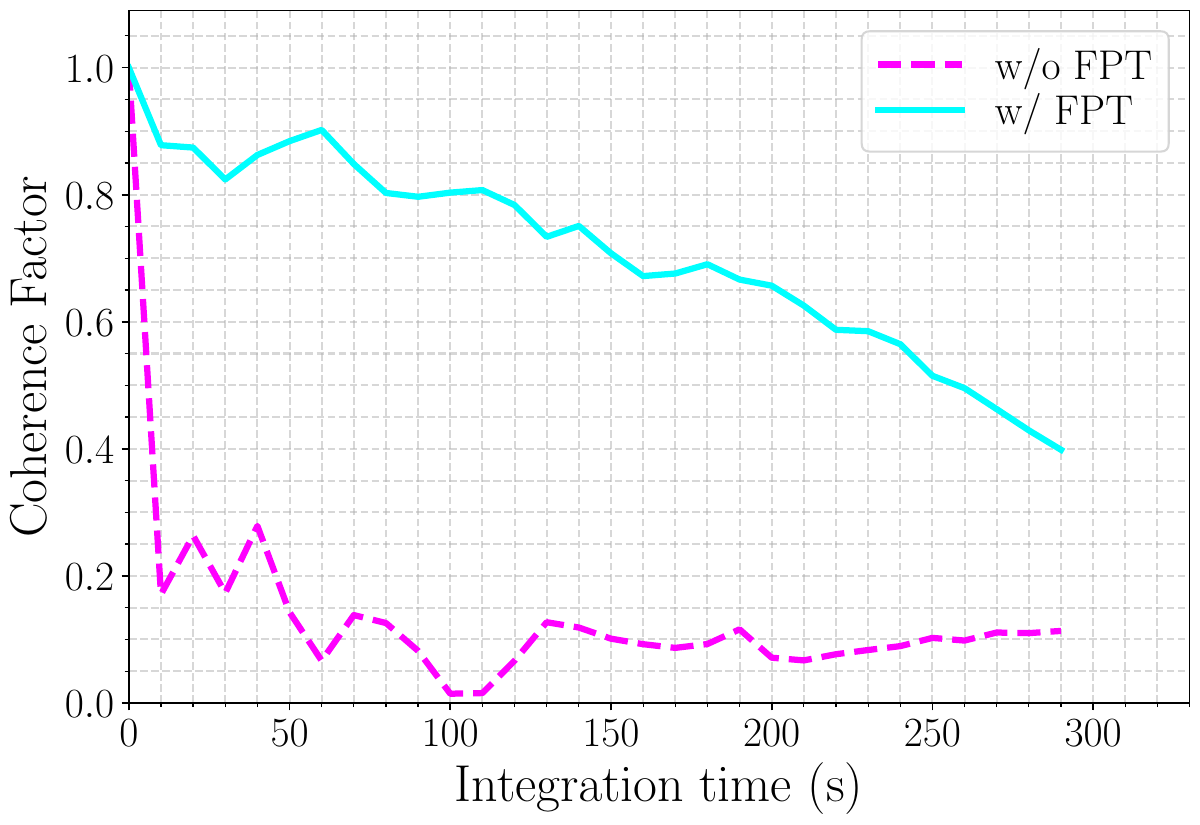}
          \caption{Same as \autoref{fig:coherence} but for the PKS\,B0420-014 scans at 00:30:00 (top) and 00:45:00 (bottom), respectively.}
             \label{fig:coherence_0420}
       \end{figure}

    In~\autoref{fig:coherence}, we show the coherence analysis on the CTA\,102 scan at 23:07:30 UT. In Figures~\ref{fig:coherence_cta_extra}--\ref{fig:coherence_0420}, we show the same results on the other scans of CTA\,102, and the scans of 3C\,454.3, and PKS\,B0420-014. The first scan of 3C\,454.3 is excluded due to a limited number of data points (see \autoref{fig:vplot}).
    In almost all scans, the coherence factor for the data after FPT remains higher than the counterpart before applying FPT. At a timescale of 1 minute, the coherence factor after FPT scales around 0.8–0.9, while the values before FPT show more distinct values, which could be related to the turbulent nature of the troposphere, but on most scans, the factor without FPT falls below 0.5 in 2 minutes or shorter. At a 5-minute timescale, the coherence factor after FPT ranges between 0.5 and 0.8, in contrast to 0.1–0.4 without FPT.

    %


\end{appendix}

\end{document}